\documentclass{article}

\usepackage{arxiv}

\usepackage{amsmath}
\usepackage{amssymb}
\usepackage{enumitem}
\usepackage{multirow}

\usepackage{algorithmic}
\usepackage{graphicx}
\usepackage{textcomp}

\usepackage{xcolor}

\usepackage[utf8]{inputenc} 
\usepackage[T1]{fontenc}    
\usepackage{hyperref}       
\usepackage{url}            
\usepackage{booktabs}       
\usepackage{amsfonts}       
\usepackage{nicefrac}       
\usepackage{microtype}      
\usepackage{lipsum}
\usepackage{graphicx}
\graphicspath{ {./images/} }

\title{Quantifying IIoT Sensor Node Criticality by Fusing its Data Criticality and Security Vulnerability}

\author{
 Sachin K. Sen \\
  School of Computing, Electrical and Applied Technology\\ Unitec Institute of Technology\\ Auckland, New Zealand. \\
  \texttt{ssen@unitec.ac.nz} \\
   \And
 Gour C. Karmakar \\
  Institute of Innovation, Science and Sustainability\\ Federation University Australia\\ Ballarat, VIC 3350, Australia. \\
  \texttt{gour.karmakar@federation} \\
  \And
 Shaoning Pang \\
  Institute of Innovation, Science and Sustainability\\ Federation University Australia\\ Ballarat, VIC 3350, Australia. \\
  \texttt{p.pang@federation.edu.au} \\
}

\begin{document}
\maketitle
\begin{abstract}
The integration of the Industrial Internet of Things (IIoT) into manufacturing has transformed industrial operations by optimising production management and ensuring product quality through smart industrial sensors that regulate processes based on real-time data. However, these sensor nodes are highly vulnerable to cyber threats, posing significant security risks that compromise their reliability and integrity. While existing research explores cybersecurity vulnerabilities and cyberattack-based methods for ranking critical nodes, some studies assess node criticality based on the impact of sensor data on product quality. However, a comprehensive approach that integrates both data criticality and cybersecurity vulnerability remains unexplored. To bridge this gap, this study introduces a novel framework that evaluates IIoT sensor node criticality by leveraging Dempster–Shafer (D-S) theory to fuse data criticality and cybersecurity vulnerabilities. The proposed method is validated using a dataset from red wine production, demonstrating its effectiveness in ranking sensor nodes based on both factors. The results show that criticality rankings based on security vulnerability scores computed using CVSS version 4.0 differ significantly from those obtained with CVSS version 3.1, highlighting the influence of enhanced vulnerability assessment methodologies. While initially applied to wine manufacturing, this framework is adaptable to broader industrial applications with minimal modifications, offering a robust approach to securing IIoT-enabled production systems.
\end{abstract}


\section{Introduction}
{T}{he} integration of smart sensors within the Industrial Internet of Things (IIoT) has transformed manufacturing by enabling real-time monitoring, process optimisation, and predictive maintenance, significantly enhancing product quality control and operational efficiency. However, ensuring the reliability, security, and effectiveness of IIoT sensor nodes remains a critical challenge. Assessing the importance of these nodes requires evaluating both the significance of their data and the associated security vulnerabilities. A combined approach integrating data-driven analysis with security assessments strengthens IIoT system resilience, ensuring safer industrial operations.

IIoT sensor hardware is deployed within manufacturing environments through wired or wireless installations, where it collects and transmits critical data via electrical signals processed by dedicated sensor software \cite{delgado2022survey}. These sensors play a crucial role in digital manufacturing systems by gathering real-time information, supporting informed decision-making, and enhancing productivity \cite{joshi2016restoring}. However, in IIoT-integrated manufacturing, a single compromised element can disrupt the entire quality assurance process. This is particularly evident in digital wine production, where key parameters such as sulfur dioxide levels, pH values, acidity, and alcohol content determine product quality. Any compromise in these elements can lead to significant deterioration, making security a paramount concern. Addressing this issue requires comprehensive security risk assessments, including an evaluation of vulnerabilities in IIoT devices. Several existing methods leverage cyberattacks such as False Data Injection (FDI) \cite{sawant2023evaluating}, Denial of Service (DoS) \cite{manjula2024enhancing}, and cybersecurity vulnerability analysis \cite{sen2024assessment} to rank critical nodes and implement protective measures. However, these methods are either attack-specific or focus solely on sensor cybersecurity vulnerabilities without considering the criticality of sensor data.

Beyond security concerns, IIoT sensor nodes also play a fundamental role in maintaining product quality in digital manufacturing systems. In sectors such as digital wine production \cite{jagatheesaperumal2021duo}\cite{corallo2021cybersecurity}, assessing node criticality based on data significance is crucial. Prior research \cite{sen2023critical} has examined sensor node criticality by evaluating the impact of data on product quality; however, it does not account for cybersecurity threats. To ensure IIoT sensor resilience, a holistic assessment is required—one that integrates both security vulnerabilities and data criticality. This dual-focus approach is essential for safeguarding IIoT-enabled production environments while maintaining and improving product quality.

However, current research does not offer methods that combine critical sensor data with security vulnerabilities to determine the overall criticality of smart IIoT sensors. The research work presented in this paper addresses this gap by fusing sensor data criticality and its security vulnerabilities using Dempster-Shafer's Evidence Theory to quantify IIoT sensor node criticality. The key contributions of our research project in addressing such grave cybersecurity concerns in IIoT-enabled digital manufacturing are as follows:

\begin{enumerate}
	
	\item For the first time, we introduce an approach to measure IIoT sensor node criticality by integrating data criticality and cybersecurity vulnerabilities using Dempster-Shafer (D-S) theory. By fusing sensor security vulnerabilities and data criticality comprehensively, our proposed assessment framework offers a robust method for evaluating IIoT sensor node criticality, potentially having significant impacts on digital manufacturing processes.
	
	\item We evaluate the fused results by calculating the probability of critical IIoT sensor nodes using $Belief$ and $Plausibility$ scores. We then rank these criticality scores and compare them with pre-fusion rankings based on sensor data criticality and security vulnerabilities. These assessments were conducted using data from the production of red wine.
\end{enumerate}

The findings of this research are expected to play a crucial role in safeguarding the future of digital manufacturing against the constantly evolving cybersecurity challenges within the evolving landscape of IIoT-enabled industries.

The subsequent sections of this report are structured as follows: Section \ref{sec:cyber-security-challenges} delves into the cybersecurity challenges inherent in IIoT-incorporated digital systems, covering critical assets, the assessment of IIoT device vulnerabilities, and the evaluation of IIoT sensor node criticalities. Section \ref{sec:literature-review} provides a comprehensive review of the current literature, covering a wide range of topics, including security vulnerabilities in IIoT sensor hardware, software, and applications, as well as the critical aspects of sensor-generated data. Section \ref{sec:approach-and-methodology} elucidates the proposed approach and methodology employed in conducting the research. Specifically, it explains how the criticality of IIoT sensor nodes has been quantified. In Section \ref{sec:results-and-discussion}, the focus shifts to the presentation and discussion of results, summarising findings and analysing the obtained results. Finally, Section \ref{sec:conclusion} brings the research article to a close with a concise discussion of research outcomes, contributions, and suggestions for future research directions.

\section{Cybersecurity Challenges in IIoT-Enabled Digital Systems}
\label{sec:cyber-security-challenges}

The defence plan against the cybersecurity risks and adversaries in IIoT-enabled digital technologies is necessary to assess the criticality of all assets. This includes critical information and data, as well as critical sensors that capture vital data, and any other system or device integrated with sensors for monitoring significant data. We also need to assess asset vulnerabilities to protect assets from potential cybersecurity breaches and minimise the impact of any breach that occurs. We need to analyse and assess possible risks against vulnerable assets, i.e., critical data and systems, which are valuable to the IIoT-enabled digital systems for delivering quality output by managing the smart production process and ensuring product quality.

\subsection{Assessment of Critical Assets in the IIoT-Enabled Digital Systems}

Assessing critical assets within IIoT-enabled digital systems is essential for determining the security measures to protect them from potential threats. To fully leverage the benefits of digital technologies in operations and production management, it's imperative to thoroughly assess critical assets, encompassing data, information, and physical assets like IoT devices and equipment involved in the systems and processes. The criticality of system components in production and manufacturing includes any disruptions that affect product quality control and productivity. According to McFarlane and Jeremy \cite{mcfarlane1999assessing}, a production system's responsiveness involves addressing disturbances and unexpected internal or external events that impact production goals. Critical asset assessment involves identifying and evaluating assets relevant to the industry's system and operations, understanding how these assets support digital industrial systems, their criticality within different system areas, and the importance of the systems they support or operate within.

Categorising assets within digital industrial systems, such as devices, information, and data, based on their importance to production management and product quality assurance, is essential for securing IIoT-integrated digital industrial manufacturing. Assets can be classified as vital, performing core functionalities, peripheral, or general, depending on their significance in the digital industrial process. Threat actors aim to compromise systems by exploiting vulnerabilities through various attack vectors. Identifying and analysing assets is crucial to assessing their criticality, which significantly impacts industrial processes. Based on the criticality and significance of assets, appropriate security controls can be implemented to address cybersecurity challenges. Threat actors compromise critical assets using vulnerabilities stemming from flaws and weaknesses associated with IIoT devices and their generated data assets. 

A rigorous process of data identification, analysis, and strategic planning is necessary to enhance the cybersecurity posture of IIoT-enabled digital manufacturing systems. Critical data support is needed in digital systems to operate and maintain sensitive and business-critical assets. Therefore, the assessment of critical data enables the process of securing critical data, sensitive equipment, and business-critical systems. The systems where critical assets reside, as well as the assets or systems upon which sensitive and business-critical components depend, are assessed for the smooth operation of the digital system \cite{isbell2019development}. Assessing critical assets requires identifying the associated risks; classifying cybersecurity risks allows digital industries to understand the exposed vulnerabilities and related risks that affect assets the most \cite{skrodelis2021cyber}. In cybersecurity, a key area is the protection of national critical infrastructures; critical infrastructures include digital manufacturing, power grids, water \& energy infrastructures, food \& beverages, and many others exposed to the cyber-world. Therefore, critical infrastructure protection encompasses all kinds of cybersecurity threats. To define the security requirements for digital systems, assessing critical assets in critical infrastructure and analysing the necessity of asset security, as well as their corresponding security goals, is proposed in \cite{mead2022critical}. 

While the technological revolution presents significant opportunities for the manufacturing sector, the integration of interconnected technologies also brings potential vulnerabilities, posing a considerable cybersecurity challenge in safeguarding against cyber threats within this digitalised landscape \cite{zhang2022advancements}. IIoT sensor nodes serve as a crucial backbone in IIoT-enabled digital manufacturing, introducing numerous security risks, including data breaches, unauthorised access, and sophisticated cyber attacks. Threat actors exploit vulnerabilities in IIoT devices and leverage various attack vectors, resulting in losses to critical assets such as essential data used for decision-making and production management. 

As industries embrace the IIoT and digitisation, safeguarding critical assets becomes essential. In IIoT-enabled digital systems, the convergence of operational technology (OT) and information technology (IT) brings both challenges and opportunities for assessing and protecting these vital assets. Conducting a thorough assessment of critical assets is crucial to guarantee the reliability, safety, and security of industrial processes and data.

\subsection{Vulnerability Assessment in the IIoT-Enabled Digital Systems}

Due to the extensive interconnectivity characteristic of IIoT, the attack surface is significantly larger than that of traditional industrial control systems. The IIoT-enabled digital industrial system has improved production processes and system efficiencies, but it has also created vulnerabilities and increased cybersecurity risks due to its exposed connectivity to the cyber world. In IIOT-enabled digital systems, multiple vulnerable areas are associated, including hardware \& software, application vulnerability, data vulnerability, IoT devices like sensor node vulnerability and many more with system flaws or weaknesses. Assessment of digital systems vulnerabilities, the identification of system weaknesses \& flaws towards decision-making about cybersecurity measures, and defining security policy to protect against potential threats and attack possibilities. A system vulnerability level and security risk assessment method have been provided in \cite{arat2023attack} that considers types of attack paths using system cybersecurity threats and vulnerability thresholds. The authors in this research have proposed a new methodology for vulnerability and risk assessment based on directed graph theory, which represents both indirect and direct vulnerabilities present in the system. An improved vulnerability assessment method has been introduced in \cite{wang2018vulnerability}, where the authors considered a solution of quantifying the degree of low attack paths and finding the complex path in the IIoT-enabled system using the maximum flow of attack and attack graph. 

Various forms of vulnerabilities in IIoT control systems contribute to expanding the attack surface within integrated IIoT technologies. To assess cybersecurity in Industrial Control Systems (ICS) of the System Control and Data Acquisition (SCADA) systems, the authors in \cite{upadhyay2020scada} proposed a vulnerability assessment method based on field knowledge acquired from real incident responses. They considered the field incident response for SCADA field devices and controllers, which were manufactured at a time when cyber-attacks were not a major concern. Therefore, there is a possibility that the system may be exposed to malware attacks and easily impacted by high network traffic. 

\subsection{Assessment of IIoT Sensor Node Criticality}

Digital systems are being advantaged by IIoT-enabled technologies, but due to their exposure to the cyber world, the evolution of vulnerabilities is subsequently causing an increased volume of cybersecurity threats. Cyber threats are assessed considering any potential danger to the system and occur due to the system vulnerabilities. Cybersecurity threats are increasing due to the wide implementation of IIoT-enabled digital systems. Threat assessment in IIoT-enabled systems can meet the IIoT system's confidentiality and integrity requirements by sharing threat intelligence, thereby building an active defence system \cite{zhang2022tiia}. The authors in \cite{tariq2020context} have proposed a qualitative threat assessment method for quantifying the assessment and remediation of systems that rely on big data, data analytics, and robotics.

Integrating IIoT with control systems increases the number of endpoints, encompassing IIoT sensors and devices, each of which could potentially serve as a gateway for cyber attackers to access the system \cite{saraIndustrialIot2023}. The proliferation of endpoints is inevitable, and with each new endpoint, the attack surface and likelihood of attempted attacks rise. Hence, it becomes imperative to identify the critical functionalities and subsequent security vulnerabilities of devices like IIoT sensors as their number increases. Exploring security vulnerabilities and threats across different IIoT subsystems involves a functional approach, including identifying security threats and vulnerabilities, as well as assessing the criticality of IIoT devices, such as sensor nodes \cite{hoffman2019industrial}.

The dynamic and interconnected nature of IIoT environments presents unique cybersecurity challenges in identifying and assessing critical IIoT devices, including sensors, necessitating a comprehensive approach to threat assessment. Resources such as \cite{zidkova2022threat}\cite{ryanOTIIOT2022} offer guidance on conducting cybersecurity risk assessments for IIoT devices. This includes utilising risk assessment matrices to identify potential threats and vulnerabilities specific to IIoT devices. It is crucial to understand the risks associated with IIoT devices and sensors, including their potential impact on process safety, reliability, environmental factors, and data confidentiality, integrity, and availability within integrated IIoT systems.

\section{Literature Review}
\label{sec:literature-review}

Identifying critical nodes, such as key IIoT devices and components in industrial processing and control systems, helps allocate appropriate security protocols and tools to ensure reliable production management. Critical nodes in IIoT networks are essential because they influence data collection, processing, network connectivity, decision-making, and system security. The criticality of IIoT devices and nodes, especially IIoT sensor nodes, is fundamental to the effectiveness of IIoT systems. These nodes are key to data collection, processing, and communication, making their reliability and security essential to the seamless operation of IIoT applications. IIoT sensor nodes are essential for ensuring operational efficiency, enhancing safety protocols, and protecting sensitive industrial data. These nodes monitor machinery, optimise processes, and support uninterrupted production flows \cite{sectrioIIoTSecurity2023,digiALERTIIoT2024}. They provide critical data that drive efficiency improvements and product quality while facing security vulnerabilities that threaten the integrity of industrial processes. This section reviews the literature on identifying IIoT sensor node criticality, grouping detection methods into three categories: (i) Cyber exploits, (ii) Data criticality, and (iii) Node failures and other factors.

\subsection{IIoT Node Criticality Based on Cyber Exploits}

Assessing the criticality of IIoT sensor nodes requires a thorough evaluation of cyber exploits, including cyber-attacks and security vulnerabilities, as these nodes play a crucial role in maintaining product quality and ensuring system operations. Threats such as FDI, DoS/DDoS, and Man-in-the-Middle (MiTM) attacks pose significant risks to IIoT ecosystems, potentially disrupting industrial processes and compromising data integrity. Furthermore, inherent cybersecurity vulnerabilities in IIoT devices and nodes contribute to an expanding cyber-threat landscape. A comprehensive risk assessment is essential for determining criticality levels and implementing targeted security measures. Such proactive evaluations enhance the overall cybersecurity posture, effectively mitigating risks and safeguarding IIoT devices against emerging threats. A proactive approach to dynamically assess the cyberattack risk of a network is introduced in \cite{cheimonidis2025proactive}. This approach determines the vulnerability along an attack path by combining system vulnerabilities and asset interdependencies. The use of Markov chains, Bayesian models, and exploit likelihood within a short time window makes the measure suitable for time-sensitive cyber-risk assessment. However, since this approach assesses cyberattack risks at the network level, it is not directly applicable to IIoT sensor vulnerability assessment.

The criticality of IIoT nodes is intrinsically linked to cybersecurity risks. IIoT devices, often integrated into critical industrial infrastructures (CII) and connected via wireless networks, are particularly susceptible to cyber-attacks. Devices with external network access, such as those connected via Wi-Fi, have an expanded attack surface, making them prime targets for cybercriminals \cite{FortinetIoTvulnerability2023}. Many IIoT environments perform safety-critical functions, where cyber-attacks on digital systems can have severe physical consequences. 

Sensor nodes exposed to external networks or with direct internet access face heightened risks and higher criticality levels \cite{WilliamsIIoTcybersecurity2024}. Prioritising such nodes for enhanced security measures is crucial to mitigating risks and preserving the integrity of IIoT ecosystems. A systematic evaluation of node criticality based on cyber-attacks and their subsequent impact is essential for prioritising security measures, allocating resources, and enhancing the resilience of IIoT infrastructures. Several existing research works consider cyber-attacks and their impact when assessing the criticality of IIoT sensors.

Sawant et al. \cite{sawant2023evaluating} assessed node criticality in consensus networks by considering the worst-case Induced Terminal Disagreement (ITD) resulting from FDI attacks. In such networks, node consensus is essential for collective decision-making, making ITD a valuable metric for evaluating criticality. While this approach is effective in consensus-based systems, it is not directly applicable to IIoT sensor networks. Unlike consensus networks, IIoT sensors operate independently, collecting and transmitting data directly to their gateways/edges without requiring inter-node agreement for forwarding decisions. Consequently, alternative methodologies are needed to assess the criticality of exploiting FDI attacks in IIoT environments.

Cybersecurity vulnerabilities in IIoT sensors represent a highly exploitable risk factor. In our previous work \cite{sen2024assessment}, we assessed the criticality of IIoT sensors by computing their vulnerability levels, where higher severity corresponded to greater criticality. To evaluate the security vulnerabilities of IIoT sensors, the Common Vulnerability Scoring System (CVSS) is used, which quantifies vulnerability scores based on key factors such as exploitability, impact, threat, and environmental metrics. Cyber-attacks serve as a crucial threat metric, often used as a proof of concept (PoC) to demonstrate the feasibility of an exploit.

\subsection{IIoT Sensor Node Criticality Based on Data Criticality}

\textcolor{black}{For quality-focused manufacturers, assessment of IIoT node criticality is a foundational step to identify the sensors that have the most influence on the product characteristics and process stability \cite{dao2026cyberscurity}.} Assessing the criticality of IIoT sensor nodes requires evaluating the significance of the data they generate and its impact on operational efficiency, product quality, and system reliability.

In a previous study \cite{sen2023critical}, we proposed a method for identifying critical IIoT sensor data in digital manufacturing systems. This approach systematically evaluates how variations in product components influence quality adjustments, ensuring more precise control in industrial production. It assesses the correlation between specific components and overall product quality, analysing the extent to which changes in component properties impact the final product. Additionally, it quantifies the percentage of quality variation resulting from these changes and determines the sensitivity of product quality to fluctuations in individual components. Integrating these factors enables a structured assessment of high-impact IIoT sensor data, facilitating better-informed adjustments in manufacturing processes. The study primarily focuses on evaluating IIoT sensor data based on its significance in maintaining and improving product quality, with sensor node criticality ranked accordingly.

\subsection{Criticality Based on Node Failures and Others}

Beyond leveraging the criticality of cybersecurity and IIoT sensor node data, various methodologies have been proposed to assess critical nodes, often focusing on network structure and dynamic behaviours. Yang et al. \cite{yang2020complex} introduced a method for identifying critical nodes in Industrial Control Systems (ICS) based on cascading failure analysis, which are essential for digital industrial production control. Their study highlights how the failure of a key control node can propagate through industrial systems, resulting in significant operational disruptions. However, this approach primarily considers control-node failures and does not account for failures of IIoT sensor nodes. Given that IIoT sensors play a crucial role in data acquisition and system monitoring, a more comprehensive and integrated approach is required to address ICS's unique safety and security challenges while incorporating the criticality of IIoT sensor nodes within IIoT-based infrastructures \cite{fu2023modeling}.


In addition to the impact of cascading failures, network connectivity is another key factor in assessing node criticality. However, while existing methods identify critical nodes based on connectivity, the impact of removing these nodes on overall network performance, operational efficiency, and product quality remains underexplored. The Importance of Connectivity (IC) method evaluates network resilience by analysing how node failures affect connectivity, incorporating cascading failure effects \cite{chang2022identification}. Load-based approaches, such as the load percolation model, examine how node failures propagate through a network, modelling network attacks as cascading failures in high-load environments \cite{hu2023critical}. Additionally, structural and topological approaches play a crucial role in identifying critical nodes. Degree centrality identifies nodes with the highest number of direct connections, while betweenness centrality measures how often a node acts as a bridge in the shortest paths between other nodes. K-shell decomposition determines core nodes within the network topology, whereas the Collective Influence (CI) algorithm evaluates a node’s criticality by considering both its degree and the degrees of neighbouring nodes within a specific radius \cite{yang2020critical}. Yang et al. \cite{yang2020critical} further refined this approach by proposing a method that ranks node importance based on both node degree and the number of structural holes it connects, providing a more comprehensive measure of node criticality.

High network load, particularly excessive traffic in IoT networks, serves as a key indicator of susceptibility to DoS attacks. DDoS attacks exploit traffic amplification to overwhelm and disrupt IoT nodes. A method proposed by Manjula et al. \cite{manjula2024enhancing} utilises node load as a criterion for identifying critical nodes, assessing their criticality by analysing the volume of traffic each node processes. Since nodes handling higher traffic are inherently more vulnerable to DDoS attacks, the authors applied a load-based K-shell algorithm to systematically rank node criticality, offering a structured approach to identifying high-risk nodes. However, this traffic-based assessment primarily evaluates the criticality of IIoT networking devices, such as gateways and edge nodes, while overlooking the crucial role of IIoT sensors in industrial systems.

\subsection{Motivation of the Research}

In digital manufacturing, IIoT sensor nodes are vital for ensuring operational efficiency, enforcing safety protocols, and securing sensitive industrial data. As these nodes become increasingly integral to industrial processes, their protection against cyber threats and physical failures is critical. Implementing robust cybersecurity measures is essential to prevent unauthorised access, mitigate data breaches, and maintain system reliability.

A comprehensive evaluation of IIoT sensor node criticality that accounts for both the importance of sensor data and the security vulnerabilities of the nodes is fundamental to safeguarding industrial operations. However, existing research lacks an integrated framework that considers both factors simultaneously. This gap in the literature underscores the need for a comprehensive methodology that evaluates node criticality by integrating data significance with cybersecurity risks. To address this challenge, our research aims to develop a unified approach that enhances the security and resilience of IIoT-enabled manufacturing environments.


\begin{figure*}
\centering
\includegraphics[scale=0.086] {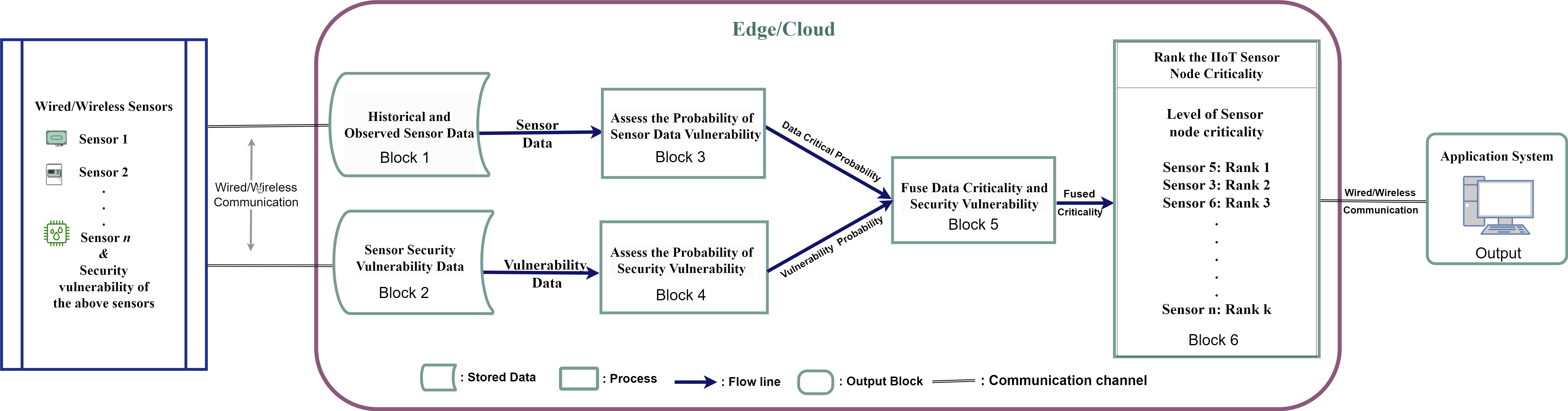}
\caption{This diagram depicts the procedure for determining the criticality of IIoT sensor nodes.}
\label{sensor-vulnerability}
\end{figure*}




\section{Proposed Approach of Quantifying IIoT Sensor Node Criticality}
\label{sec:approach-and-methodology}

This section presents a framework to assess the importance of IIoT sensor nodes in a digital manufacturing setup, considering both data criticality and sensor security vulnerabilities. The approach is demonstrated within a cutting-edge wine manufacturing context, leveraging IIoT integration and advanced data analytics for high-quality wine production.

\subsection{Overview of the Proposed Approach}

This study assesses the importance of IIoT sensor nodes in digital manufacturing systems by integrating security vulnerabilities with data criticality to mitigate potential cyber threats. Figure \ref{sensor-vulnerability} illustrates the proposed methodology for evaluating the criticality of an IIoT sensor node. The system overview highlights six blocks: Block 1, Block 2, Block 3, Block 4, Block 5, and Block 6, which are involved in the IIoT sensor node criticality calculation process. As detailed below, the methodology outlined in these blocks is utilised to achieve the research objective.

\begin{enumerate}
\item Block 1 handles IIoT sensor data collection, and Block 2 gathers sensor security vulnerability information. These data streams are then processed in Blocks 3 and 4 for probability assessments.

\item Block 3 assesses the likelihood of sensor data criticality, while Block 4 evaluates the likelihood of IIoT sensor security vulnerability.

\item Block 5 fuses the probabilities determined in Blocks 3 and 4 to quantify the criticality of an IIoT sensor node. 

\item Block 6 ranks the levels of IIoT sensor node criticality as $x$ $ \in \{1, 2, 3, \cdots k\}$, where '1' signifies the highest critical node, and '$k$' indicates the lowest critical node, Finally, the relevant application systems monitor or access the sensor node criticality for the decision-making process. The proposed method for assessing the criticality of the IIoT sensor node follows these principles:
\begin{enumerate}
	\item The higher the criticality of a sensor node, the more important its data.
	\item  Nodes with higher criticality levels are more likely to be compromised.
\end{enumerate}

\item The application systems employ the rankings of node criticalities.

\item All the processes required for Blocks 1-6 are executed by either Cloud Service or Edge Systems, depending on the Manufacturing Systems in use.
\end{enumerate}

\subsection{Assessing the Probability of Sensor Data Criticality}
The likelihood of an IIoT sensor node being deemed critical depends on the importance of the data it generates. This probability is assessed by evaluating the importance of the data for decision-making or system functionality \cite{sen2023critical}. When the data produced by the sensor is considered critical, the probability of the sensor node being critical rises due to the potential significant repercussions on the product quality of a manufacturing system.

	\begin{align}
	\label{eqn:probability-data-critical}
 m _{i1}(\zeta)=1-\frac{\mathcal{R}_{i1}}{n} \end{align}
	where,
	\begin{center}\begin{tabular}{ccl}\
			$\mathcal{R}_{i1}$ & = & Data criticality rank of the $i^{th}$ sensor.\\	
			$n$ & = & Number of sensors\\ 
			$m _{i1}(\zeta)$ & = & Probability of $i^{th}$ sensor being critical\\
			&  & considering its data criticality i.e., second \\
			&  & subscript 1 represents data criticality. \\
			$\zeta$ & = & Observed data/vulnerability level being\\
			&   & critical.\\
	\end{tabular}\end{center}

\subsection{Assessing the Probability of Sensor Security vulnerability}

Several techniques are available in the literature, such as CVSS versions 4.0 and 3.1, to assess the cybersecurity vulnerability level of an IIoT sensor node [reference to our second paper]. In CVSS, vulnerability scores range from 0 to 10, with higher scores indicating a greater security vulnerability. Severity levels are categorised as: 0.1 to 3.9 is labelled as "Low," 4.0 to 6.9 as "Medium," 7.0 to 8.9 as "High," and 9.0 to 10.0 as "Critical". By scrutinising these vulnerabilities, we can gauge the probability of a sensor node being considered critical regarding its potential impact on both system security and functionality.\\

The probability of a sensor node being critical based on cybersecurity vulnerability can be defined as,
		
	\begin{align}
	\label{eqn:probability-security-vuln}
 m _{i2}(\zeta)=1-\frac{\mathcal{R} _{i2}}{n} \end{align}
	where,
	\begin{center}\begin{tabular}{ccl}\
			$\mathcal{R}_{i2}$ & = & Security vulnerability rank of the $i^{th}$ sensor.\\	
			$m _{i2}(\zeta)$ & = & Probability of the $i^{th}$ sensor being\\
			&  & critical for cybersecurity.\\
	\end{tabular}\end{center}			
\subsection{Fusing Data Criticality and Cybersecurity Vulnerability}
Quite a few methods use subjective logic for decision fusion, such as Bayesian theory, rule-based in the reference systems, and the D-S Theory. For the fusion, we choose the D-S Theory among those methods in the literature because the D-S Theory, known as the evidence theory of belief functions, supports different propositions using temporal data and is based on the generalised Bayesian theory \cite{chowdhury2020trustworthiness}. The popularity of the D-S Theory has grown significantly in various industrial applications due to its commutative and associative nature. As this theory is widely employed in fault detection, pattern recognition, decision-making, statistical classification, medical diagnosis, and reliability analysis, it will be an effective method for fusing fuse data criticality and cybersecurity vulnerability.   

\subsubsection{Dempster-Shafer Evidence Theory}

The D-S Theory consists of three salient functions - ($a$) basic probability assignment (bpa), ($b$) the Belief function (Bel), and (c) the Plausibility function (Pl), which are propositioned as follows: 

\textbf{Proposition 1:} If we consider a mutually exclusive finite set that belongs to the discernment frame $X$, the elements of $X$ are in the power set $P(X)$. The $bpa$ assigns a mass value, denoted by $m$, which  maps an element of the power set $P(X)$ to an interval [0, 1], where $bpa$ of the null-set ($\emptyset$), $m(\emptyset)=0$. The sum of all the $bpa$'s of all the elemental sets of the power set will be equal to 1. The $bpa$ of a given set $A$, $m(A)$ is the $bpa$ value of only the set $A$, not for its subsets. The $bpa$ of any subset of $A$ will contain the independent value. The $bpa$ can now be propositioned as,

\begin{equation} 
	\label{eqn:bpa-0-1}
m: P(X)\rightarrow [0,1]
\end{equation}

\begin{equation}
	\label{eqn:bpa-1}
\text{and,} \sum _{A \in P{(X)}} m(A) = 1
\end{equation}

\textbf{Proposition 2:} The "Belief" and "Plausibility" are the $lower$ and $upper$ bounds of an interval that $bpa$ defines and bounded by the two continuous measurements. The "Belief" of a set "A" is the summation of all its subsets (say, "B"). The Belief function ($Bel$) is propositioned as,
	
\begin{equation} 
	\label{eqn:Bel-A}
{\mathrm{ Bel}}(A) = \sum _{B|B\subseteq A} m(B).
\end{equation}

where,
\begin{center}\begin{tabular}{ccl}\
		$Bel(A)$ & = & the lower bound of set $A$, $(A \in P(X))$ \\
		$B$ & = & the subset of $A$, $B \subseteq A$ \\
		$m(B)$ & = & the $bpa$ of $B$ 
\end{tabular}\end{center}		

\textbf{Proposition 3:} The $Plausibility$ ($upper$ bound of the interval) is defined as the sum of all the $bpa$ values of the sets $B$ that intersect the set $A$ ($B \cap A$). The $Plausibility$ function ($Pl$) can be propositioned as,

\begin{equation}
	\label{eqn:Pl-A-1}
{\mathrm{ Pl}}(A) = \sum _{B|B \cap A \neq \emptyset } m(B)
\end{equation}

where,
\begin{center}\begin{tabular}{ccl}\
		$Pl(A)$ & = & the $Plausibility$ of $A$ \\
		$B$ & = & a non-null set that intersects $A$ ($B \cap A  \neq \emptyset$) \\
\end{tabular}\end{center}

\textbf{Proposition 4:} The $Belief$ and the $Plausibility$ measures are derived from each other if the sum of all basic assignments equals 1. The derivation is as follows:

\begin{equation}
	\label{eqn:Pl-A-2}
{\mathrm{Pl}}(A) = 1 - {\mathrm{ Bel}}(\bar{A})
\end{equation}

where, $\bar{A}$ is the classical complement of $A$.

\subsubsection{Dempster's Rule of Combination}

The Dempster's rule \cite{shafer1976mathematical} can combine multiple independent evidences (Belief functions) with their $bpa$ (m) values and use the conjunctive operator "AND". Belief function for the two $bpa$ values of $m_{i1}(\zeta)$ and $m_{i2}(\zeta)$ are computed as below:

\begin{equation}
	\label{eqn:DEmpsterCombination}
Bel_{i}(\zeta) = \frac {1}{1-k} \times \sum^{} _{m_{i1}(\zeta) \cap m_{i2}(\zeta)} m_{i1}(\zeta)m_{i2}(\zeta)
\end{equation}

for $m_{ij}$($\zeta) \neq \emptyset$, where, $m_{ij}(\emptyset)$ is known as null value of the $bpa$, which is equal to $0$.

and,

\begin{equation} 
	\label{eqn:Dempster-K}
k = \sum _{m_{i1}(\zeta) \cap m_{i2}(\zeta)=\varnothing} m_{i1}(\zeta)m_{i2}(\zeta)
\end{equation}

Here, $k$  is the fundamental conflict probability mass, and $(1-k)$ is the normalisation factor in Dempster's Rule. $K$ is computed by summing up the $bpa$ values of all sets where the intersection is null. 

\subsubsection{Frame of Discernment}

For the two evidential sources, such as data criticality and cybersecurity vulnerability, the discernment frame $X$ introduced in Proposition 1 is defined as,

\begin{align}
	\label{eqn:DiscernmentFrame}
X = ({\zeta, \bar{\zeta}}) 
\end{align}

where $\zeta$ and $\bar{\zeta}$ represent observed data/vulnerability level being critical and not being critical, respectively.

According to Shannon's Information Theory, the more predictable an event is, the less uncertainty it holds, so acquiring information reduces this uncertainty \cite{dolors2024entropy}. Uncertainty is maximised when the mass value of $m_{ij} (\zeta) = 0.5$; any deviation from 0.5, in either direction, decreases uncertainty. 

Following the principles of Shannon's Information Entropy Theory, the uncertainty ${m_{ij}}(\zeta \vee \bar{\zeta})$ is calculated as,

\begin{equation}
	\label{eqn:ProbUncertaintyDataCritical} 
\begin{split} {m_{ij}}(\zeta \vee \bar{\zeta}) = min(-m _{i j}(\zeta)\text{log}_2m _{i j}(\zeta) \\ -(1-m _{i j}(\zeta)) \text{log}_2 (1-m _{ij}(\zeta)), 1-m _{ij}(\zeta))
\end{split}
\end{equation}
where, $j \in \{1, 2\}$, $j = 1$ for data criticality and $j = 2$ for cybersecurity vulnerability.

Considering probabilistic measures of the mass and uncertainty values, we can define $m_{ij}(\bar{\zeta})$ as follows:

\begin{align}
	\label{eqn:UncertaintyDataCritical}
m_{ij}(\bar{\zeta})=1-m _{ij}(\zeta)-m _{ij}({\zeta \vee \bar{\zeta}}).
\end{align}

In some cases, $m_{ij}(\zeta \vee \bar{\zeta})+ m _{ij}(\zeta)>1$. When this occurs, the value of $m_{ij}(\bar{\zeta})$ will be negative. To prevent this, we use the min() function in Eq.  \ref{eqn:ProbUncertaintyDataCritical}. By applying the values of $m _{i j}(\zeta)$, ${m_{ij}}(\zeta \vee \bar{\zeta})$, and $m_{ij}(\bar{\zeta})$ to Eq. \ref{eqn:DEmpsterCombination}, we can calculate the lower limit of the $i^{th}$ sensor node being critical, $Bel_i(\zeta)$, and the upper limit, $Pl_i(\zeta) = 1 - Bel_{i}(\bar{\zeta})$ (refer to Eq. \ref{eqn:Pl-A-2}).

\subsection{Limitations and Biases}

We computed the criticalities and security vulnerabilities of sensor-generated data for a limited selection of sensors to monitor the wine production process. Security vulnerabilities were assessed using the CVSS method, with exploitable and impact metric values defined based on publicly available information. Additionally, data criticalities were determined using various statistical methods and predictive models were developed that considered multiple factors. As a result, the project may be subject to certain limitations and biases, including:\\

\begin{enumerate}
        \item The project endeavours to quantify sensor node criticality by integrating data criticality and security vulnerability. However, the subjective nature of defining and measuring these factors leaves room for human error and bias.
        \item The D-S Theory, employed in the analysis, comes with its own set of assumptions and limitations that could impact the accuracy and reliability of the project's findings.
        
        \item Generalising the project's results to all IIoT sensor nodes may not be feasible, as criticality and security vulnerability can vary significantly depending on the specific application and environmental factors.
        
        \item The process of assigning weights to different factors, such as data criticality and security vulnerability, involves subjective judgment, which can potentially introduce bias into the analysis. Stakeholders with differing priorities may produce varying assessments of sensor node criticality.
        
        \item The validity and applicability of the metrics utilised to measure security vulnerability in IIoT sensor nodes may not be fully accurate as we considered publicly available information in our computation. The lack of accuracy in our assumption of these metrics might not adequately capture potential threats and vulnerabilities, and the overall criticality assessment may be compromised.
\end{enumerate}

\section{Results and Discussion}
\label{sec:results-and-discussion}
In this research, we effectively utilise a new method of measuring the importance of IIoT sensor nodes to protect IIoT-incorporated digital manufacturing. By combining the criticality of sensor data and sensor security vulnerability metrics, we assess the potential risks associated with these industrial nodes. The findings offer a detailed insight into the impact of IIoT sensor vulnerabilities on overall product quality in a digital manufacturing environment.

\subsection{Case Scenario}

We examine a state-of-the-art wine manufacturing facility that employs advanced IIoT sensor technologies to enhance wine quality and customisation. Here, we assume that the wine manufacturing process utilises a variety of IIoT sensors to monitor crucial parameters, including acidity levels (concentration of acids, volatile acidity, citric acid), sugar levels, chloride level (controls $NaCl$ in wine), sulphate, $SO_2$, pH, the density of wine, and alcohol levels. These sensors play a vital role in adapting and adjusting these parameters during the wine production process, ensuring the maintenance of the wine's colour and flavour profile for producing high-quality wine. These sensors' detailed insights enable precise control over the wine's composition.

In this scenario, the challenge is consistently upholding wine quality while securing the digital wine manufacturing process. Maintaining the quality of wine involves ensuring the consistency of sensor-generated data and the security of the sensors. Key considerations include maintaining data confidentiality, integrity, and availability, minimising threat levels, impacts, and exploitability, and defending against potential attack vectors and vulnerabilities. An attack on the communication of IIoT sensors could lead to manipulated data, impacting decisions at the fermentation duration, blending ratios, or bottling timing. Compromised sensor data has the potential to alter the wine's flavour, colour, or aroma profiles, ultimately affecting the overall quality of the final product.

\begin{table*}[]
	\caption{The probabilities of the mass functions and corresponding uncertainties associated with the IIoT sensor-generated data criticality and cybersecurity vulnerabilities within the IIoT incorporated digital manufacturing.}
	\resizebox{\textwidth}{!}{%
\bgroup
\def\arraystretch{1.7}%
\begin{tabular}{lllllllllllll}
\hline
\multirow{2}{*}{Sensor ID} &
  \multirow{2}{*}{} &
  \multicolumn{3}{l}{Probabilities for Sensor Data Criticality} &
  \multirow{2}{*}{} &
  \multicolumn{3}{l}{\begin{tabular}[c]{@{}l@{}}Probabilities of Sensor Security Vulnerabilities\\[-1.5ex] (Computed with CVSS 4.0)\end{tabular}} &
  \multirow{2}{*}{} &
  \multicolumn{3}{l}{\begin{tabular}[c]{@{}l@{}}Probabilities of Sensor Security Vulnerabilities\\[-1.5ex] (Computed with CVSS 3.1)\end{tabular}} \\ \cline{3-5} \cline{7-9} \cline{11-13} 
 &
   &
  \multicolumn{1}{l}{$m_{i1}(\zeta)$} &
  \multicolumn{1}{l}{$m _{i1}({\bar{\zeta}})$} &
  $m _{i1}({\zeta \vee \bar{\zeta}})$ &
   &
  \multicolumn{1}{l}{$m_{i2}(\zeta^{v^{4.0}})^{\dagger}$} &
  \multicolumn{1}{l}{$m _{i2}({\bar{\zeta}^{v^{4.0}}})$} &
  $m _{i2}({\zeta^{v^{4.0}} \vee \bar{\zeta}^{v^{4.0}}})$ &
   &
  \multicolumn{1}{l}{$m_{i2}(\zeta^{v^{3.1}})^{\dagger}$} &
  \multicolumn{1}{l}{$m _{i2}({\bar{\zeta}^{v^{3.1}}})$} &
  $m _{i2}({\zeta}^{v^{3.1}} \vee \bar{\zeta}^{v^{3.1}})$ \\ \hline
S01 &
  \multirow{11}{*}{} &
  0.8182 &
  0.0000 &
  0.1818 &
  \multirow{11}{*}{} &
  0.8182 &
  0.0000 &
  0.1818 &
  \multirow{11}{*}{} &
  0.8182 &
  0.0000 &
  0.1818 \\[-1ex]
S02 &
   &
  0.6364 &
  0.0789 &
  0.2847 &
   &
  0.7273 &
  0.0182 &
  0.2545 &
   &
  0.6364 &
  0.0789 &
  0.2847 \\[-1ex]
S03 &
   &
  0.0909 &
  0.7768 &
  0.1323 &
   &
  0.8182 &
  0.0000 &
  0.1818 &
   &
  0.8182 &
  0.0000 &
  0.1818 \\[-1ex]
S04 &
   &
  0.7273 &
  0.0182 &
  0.2545 &
   &
  0.8182 &
  0.0000 &
  0.1818 &
   &
  0.8182 &
  0.0000 &
  0.1818 \\[-1ex]
S05 &
   &
  0.3636 &
  0.3517 &
  0.2847 &
   &
  0.8182 &
  0.0000 &
  0.1818 &
   &
  0.8182 &
  0.0000 &
  0.1818 \\[-1ex]
S06 &
   &
  0.2727 &
  0.4728 &
  0.2545 &
   &
  0.8182 &
  0.0000 &
  0.1818 &
   &
  0.8182 &
  0.0000 &
  0.1818 \\[-1ex]
S07 &
   &
  0.9091 &
  0.0000 &
  0.0909 &
   &
  0.8182 &
  0.0000 &
  0.1818 &
   &
  0.8182 &
  0.0000 &
  0.1818 \\[-1ex]
S08 &
   &
  0.0909 &
  0.7768 &
  0.1323 &
   &
  0.9091 &
  0.0000 &
  0.0909 &
   &
  0.9091 &
  0.0000 &
  0.0909 \\[-1ex]
S09 &
   &
  0.1818 &
  0.6123 &
  0.2059 &
   &
  0.9091 &
  0.0000 &
  0.0909 &
   &
  0.9091 &
  0.0000 &
  0.0909 \\[-1ex]
S10 &
   &
  0.4545 &
  0.2463 &
  0.2992 &
   &
  0.7273 &
  0.0182 &
  0.2545 &
   &
  0.6364 &
  0.0789 &
  0.2847 \\[-1ex]
S11 &
   &
  0.5455 &
  0.1553 &
  0.2992 &
   &
  0.8182 &
  0.0000 &
  0.1818 &
   &
  0.8182 &
  0.0000 &
  0.1818 \\ \hline
\multicolumn{10}{p{341pt}}{\small $\dagger\zeta^{v^{4.0}} \;and\;\dagger\zeta^{v^{3.1}}$ indicate the security vulnerability using CVSS4.0 and CVSS3.1 respectively.} \\
\end{tabular}
\egroup
	}
	\label{Prob-and-Uncertainty}
\end{table*}

\subsection{Dataset - Evidences for Fusion}
We implement the data criticality detection framework proposed in our prior research \cite{sen2023critical} within the mentioned scenario, utilising the wine production dataset.

The red wine dataset comprises 11 independent characteristics and their corresponding wine quality ratings. Wine quality varies based on changes in individual characteristics. The characteristic values represent the $i^{th}$ wine's characteristic and quality, respectively. Here, $ i \in {1, 2, 3, 4, 5, 6, 7, 8, 9, 10, 11} $, representing the index of the set {\textit{"Fixed Acidity (FA)", "Volatile Acidity (VA)", "Citric Acid (CA)", "Residual Sugar (RS)", "Chlorides (CL)", "Free Sulphur Dioxide (FSD)", "Total Sulphur Dioxide (TSD)", "Density (DT)", "pH (PH)", "Sulphates (SLP)", "Alcohol (ALC)"}}. The sensors measuring the wine characteristic values are labelled S01, S02, S03, S04, S05, S06, S07, S08, S09, S10, and S11, respectively.

\subsubsection{IIoT Sensor-Generated Critical Dataset}

We construct a predictive model using both non-machine-learning and machine-learning techniques on the wine dataset. Subsequently, we employ a key method to evaluate the disparity between predicted and reference wine quality in response to changes in wine quality values. The criteria for assessing data criticality include:

\begin{enumerate}
\item The correlation between the instantaneous data values of a wine characteristic and its estimated wine quality values.

\item The percentage change in wine quality for a specific wine characteristic.

\item The sensitivity of wine quality to changes in a wine characteristic.
\end{enumerate}

Finally, we calculate the overall data criticality levels using weighted, normalised ranking values. The criticality values for wine parameters that influence wine quality are integrated into the fusion dataset.

\subsubsection{IIoT Sensor Security Vulnerability Dataset}

To construct the additional fusion dataset that incorporates IIoT sensor security vulnerabilities, we used the same IIoT-integrated digital wine manufacturing scenario described earlier. We employed the well-established Common Vulnerability Scoring System (CVSS) methodology to compute sensor security vulnerabilities. Throughout this process, we ensured alignment with CVSS metric assessment criteria by defining exploitability, impact, and threat metric values as follows:

\begin{enumerate}

\item We established exploitability, impact, and threat metric values tailored to IIoT wine sensors, while maintaining compatibility with the CVSS methodology.

\item For assessing impact metrics using the Base Score within the wine manufacturing context, we utilised correlations between wine characteristics and their respective reference wine quality values.

\item The analysis of threat metrics is environment-specific. However, we comprehensively evaluated potential threats and associated threat metrics for IIoT wine sensors, considering vulnerability exposures within the digital wine manufacturing system architecture. This system involves sensor network communication within a shared WiFi network, where sensors communicate with the central controller (PLC - Programmable Logic Controllers) and utilise public cloud storage.
\end{enumerate}

Applying the aforementioned methods and defined CVSS metric values, along with the use of a CVSS calculator, we computed the IIoT sensor security vulnerability scores, constituting the additional fusion dataset for this research.

\subsubsection{Probabilities and Uncertainties Associated with the Evidence}

Dempster's fusion theory requires understanding the probabilities and uncertainties associated with the evidence dataset. By utilising the prepared evidence dataset and employing the relevant equations outlined in the proposed methodology section, we calculate the required probabilities and associated uncertainties for the evidence. Specifically, we use equations (Eqs. \ref{eqn:probability-data-critical} and \ref{eqn:probability-security-vuln}) to determine the probabilities of critical sensor-generated data and security vulnerabilities of the sensors. Correspondingly, we use other equations (Eqs. \ref{eqn:ProbUncertaintyDataCritical}-\ref{eqn:UncertaintyDataCritical}) to compute the uncertainties linked to these probabilities. These computations yield insights into the probabilities and uncertainties concerning sensor-generated data criticality and security vulnerabilities. The resulting probabilities and uncertainties are detailed in TABLE \ref{Prob-and-Uncertainty}.

\subsection{Sensor Node Criticality Fusing Data Criticality and Security Vulnerability}

Our proposed methodology considers both the criticality of data and the security vulnerabilities of IIoT sensor nodes, allowing us to accurately assess their overall criticality. By integrating these dimensions effectively through our fusion model, we better represent the potential risks associated with each sensor node.

We have two sets of evidence: Set-$A$, which includes (i) sensor-generated critical data and (ii) sensor security vulnerability computed using CVSS 4.0 metrics, and Set-$B$, which comprises (i) sensor-generated critical data and (ii) sensor security vulnerability computed using CVSS 3.1 metrics. The fusion method generates distinct levels of IIoT sensor node criticality from each set, which will be discussed in detail in the following sections.

\begin{table*}[]
	\caption{The prioritisation (Ranking) of IIoT sensor node criticality within IIoT-enabled digital manufacturing, determined by fusing its data criticality and cybersecurity vulnerabilities. The Prioritisation Ranking is compared with the Sensor-generated Data Criticality Ranking and Security Vulnerability Ranking based on CVSS4.0 and CVSS3.1, respectively.}
	\resizebox{\textwidth}{!}{%
\bgroup
\def\arraystretch{1.7}%
\begin{tabular}{llllllllllllllll}
\hline
\multirow{2}{*}{Sensor ID} &
   &
  \multicolumn{3}{l}{\begin{tabular}[c]{@{}l@{}}Ranking Based on Probability Scores by Fusing Data\\[-1.5ex] Criticality and Security Vulnerability using CVSS4.0\end{tabular}} &
   &
  \multicolumn{3}{l}{\begin{tabular}[c]{@{}l@{}}Ranking Based on Probability Scores by Fusing Data\\[-1.5ex] Criticality and Security Vulnerability using CVSS3.1\end{tabular}} &
   &
  \begin{tabular}[c]{@{}l@{}}Ranking Based on \\[-1.5ex] Data Criticality\end{tabular} &
   &
  \begin{tabular}[c]{@{}l@{}}Criticality Based on\\[-1.5ex] CVSS4.0 Vulnerability\end{tabular} &
   &
  \begin{tabular}[c]{@{}l@{}}Criticality Based on\\[-1.5ex] CVSS3.1 Vulnerability\end{tabular} &
   \\ \cline{3-5} \cline{7-9} \cline{11-11} \cline{13-13} \cline{15-16} 
    &  & \multicolumn{1}{c}{$R^{Bel({\zeta^{v^{4.0}}_{d}})}\ddagger$} &  & $R^{Pl({\zeta^{v^{4.0}}_{d}})}$ &  & ${R^{Bel({\zeta^{v^{3.1}}_{d}})}\ddagger}$ &  & $R^{Pl({\zeta^{v^{3.1}}_{d}})}$  &  & $R^{d}$  &  & $R^{v^{4.0}}$ &  & $R^{v^{3.1}}$ &  \\ \hline
S07 &  & 1                                                              &  & 1 &  & 1 &  & 1 &  & 10 &  & 1    &  & 3    &  \\[-1ex]
S01 &  & 2                     &  & 2 &  & 2 &  & 2 &  & 8 &  & 2   &  & 2 &  \\[-1ex]
S04 &  & 3                     &  & 3 &  & 3 &  & 3 &  & 2  &  & 5    &  & 4    &  \\[-1ex]
S11 &  & 4                     &  & 5 &  & 4 &  & 4 &  & 9  &  & 9    &  & 5    &  \\[-1ex]
S02 &  & 5                     &  & 4 &  & 5 &  & 5 &  & 5  &  & 6   &  & 6   &  \\[-1ex]
S05 &  & 6                     &  & 7 &  & 7 &  & 7 &  & 4  &  & 7    &  & 7    &  \\[-1ex]
S09 &  & 7                     &  & 6 &  & 9 &  & 6 &  & 11  &  & 11    &  & 8    &  \\[-1ex]
S10 &  & 8                     &  & 9 &  & 6 &  & 10 &  & 1  &  & 3    &  & 10   &  \\[-1ex]
S06 &  & 9                     &  & 10 &  & 8 &  & 9 &  & 3  &  & 4    &  & 11    &  \\[-1ex]
S08 &  & 10                     &  & 8 &  & 10 &  & 8 &  & 6  &  & 10    &  & 1    &  \\[-1ex]
S03 &  & 11                     &  & 11 &  & 11 &  & 11 &  & 7  &  & 8    &  & 9    &  \\ \hline
\multicolumn{16}{p{451pt}}{\small $\ddagger\zeta_{d}^{v^{4.0}} and \;\ddagger\zeta_{d}^{v^{3.1}}$ indicate the combined mass of data criticality \& security vulnerability for CVSS4.0 and CVSS3.1 respectively.} \\
\end{tabular}
\egroup
	}
	\label{Criticality-Probability}
\end{table*}

\subsubsection{Fusing Data Criticality and CVSS 4.0 Security Vulnerability}

\begin{figure}[htbp]
    \centering
    \includegraphics[width=1.0\textwidth]{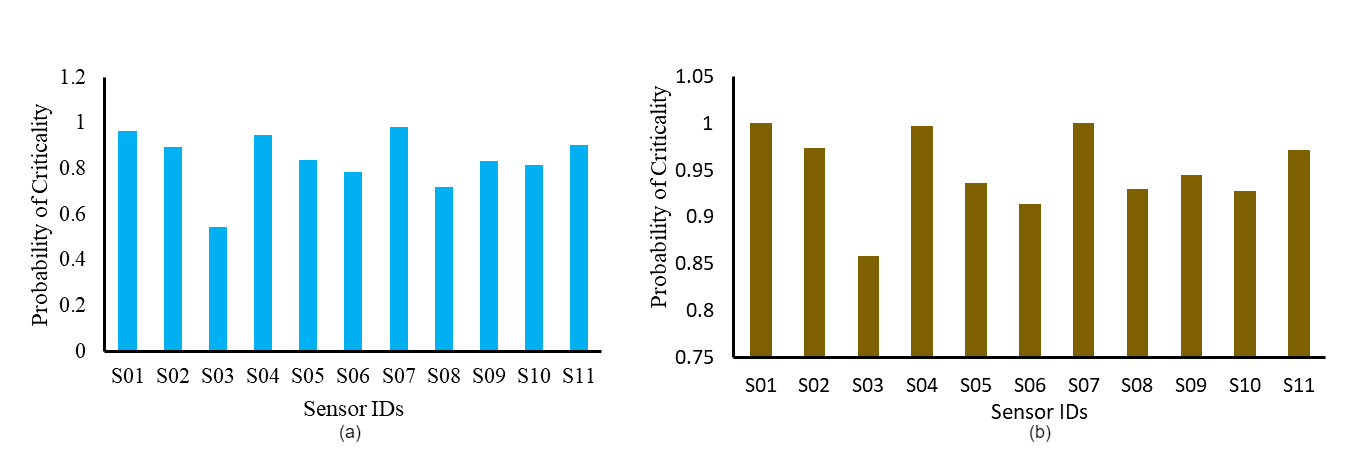} 
    \caption{The diagram illustrates the probability of critical sensor nodes based on the fused evidence from Set-$A$. Part (a) shows the lower limit of the probability, represented by belief scores, while part (b) presents the upper limit, represented by plausibility scores.}
    \label{BEL-PL-CVSS40}
\end{figure}

Combining data from Set-$A$, critical data generated by IIoT sensors, and their security vulnerabilities (calculated using the CVSS 4.0 calculator), we assess the criticality of IIoT sensor nodes. This approach provides a more accurate estimation by leveraging Dempster's rule of combination to account for the probabilities and uncertainties associated with data criticality and cybersecurity vulnerabilities. The resulting probabilities of IIoT sensor nodes being critical, represented by Belief scores (lower limits) and Plausibility scores (upper limits), are depicted in Figure \ref{BEL-PL-CVSS40}.


As shown in Figure \ref{BEL-PL-CVSS40}(a), the belief (Bel) scores of Set-$A$ across all 11 sensors range from 0.5465 to 0.9835, indicating a high level of confidence in the combined evidence. Figure \ref{BEL-PL-CVSS40}(b) illustrates that the plausibility (Pl) scores range from 0.8588 to 1.0, suggesting a high level of plausibility in the evidence. Bel (support) and Pl (believable) scores represent the lower and upper probability limits, respectively, with Bel always less than or equal to Pl. Bel and Pl reflect the degree of probabilities, where higher values indicate higher sensor criticality. Notably, sensors S02, S11, S04, S01, and S07 are particularly critical based on their Bel values (0.8934, 0.9053, 0.9496, 0.9669, 0.9835) and Pl values (0.9733, 0.9718, 0.9967, 1.0000, 1.0000), respectively. These sensors require closer attention while allocating security protocols due to their elevated risk levels compared to others.

\subsubsection{Fusing Data Criticality and CVSS 3.1 Security Vulnerability}

\begin{figure}[htbp]
    \centering
    \includegraphics[width=1.0\textwidth]{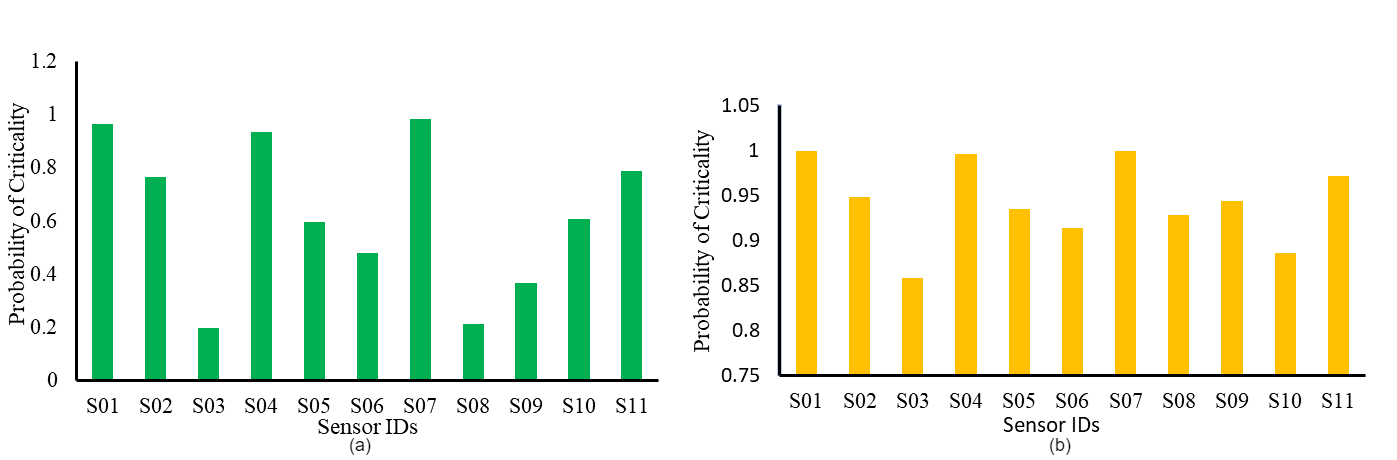} 
    \caption{This diagram illustrates the probability of critical sensor nodes based on the fused evidence from Set-$B$. Part (a) displays the lower limit of the probability, represented by belief scores, while part (b) shows the upper limit, represented by plausibility scores.}
    \label{BEL-PL-CVSS31}
\end{figure}

We also integrated evidence from Set-$B$, which includes the same IIoT sensor-generated data criticality and corresponding sensor security vulnerabilities calculated using the CVSS version 3.1 calculator. The fusion results for this dataset, showing the probabilities of IIoT sensor nodes being critical along with their $Bel$ and $Pl$ scores, are presented in Figure \ref{BEL-PL-CVSS31}.

As depicted in Figure \ref{BEL-PL-CVSS31}(a), the $Bel$ scores across all 11 sensors in Set-$B$ range from 0.1992 to 0.9834, indicating a broader range of belief in the combined evidence compared to Set-$A$. Figure \ref{BEL-PL-CVSS31}(b) shows that the $Pl$ scores range from 0.8587 to 1.0, similar to those in Set-$A$. Sensors S02, S11, S04, S01, and S07 are highlighted as more critical based on their $Bel$ values (0.7673, 0.7903, 0.9355, 0.9669, 0.9835) and $Pl$ values (0.9488, 0.9718, 0.9967, 1.0000, 1.0000) respectively. These sensors warrant closer attention for security protocol allocation due to their elevated risk levels compared to others.

\begin{figure}[htbp]
    \centering
    \includegraphics[width=1.0\textwidth]{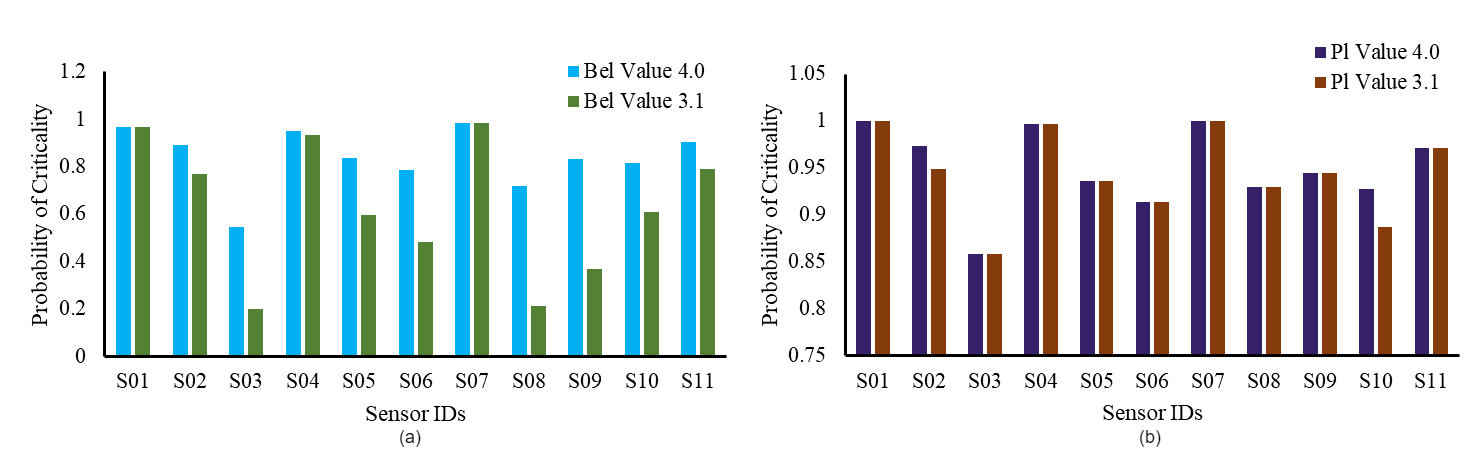} 
    \caption{This diagram provides a comparative analysis of the probabilities, represented by $Bel$ and $Pl$ scores, based on the fused results of two sets of evidence. Part (a) compares the $Bel$ scores obtained from fusing evidence in Set-$A$ with those from Set-$B$. Part (b) compares the $Pl$ scores from fusing the two sets.}
    \label{BEL-PL-COMPARECVSS40-and-31}
\end{figure}


\subsection{Comparative Analysis of the Results}

We compare the results obtained from fusing the evidence in Set-$A$ and Set-$B$. Using the computed $Bel$ (belief) and $Pl$ (plausibility) scores, we rank the criticality of IIoT sensor nodes, as presented in TABLE \ref{Criticality-Probability}. This table includes the $Bel$ and $Pl$ values derived from the fusion of both evidence sets. The table also includes pre-fusion criticality rankings based on sensor-generated data and sensor security vulnerabilities, computed using CVSS 4.0 and CVSS 3.1 metrics to provide a comprehensive comparison.

A visual comparison is also provided in Figures \ref{BEL-PL-COMPARECVSS40-and-31}. Figure \ref{BEL-PL-COMPARECVSS40-and-31}(a) offers a side-by-side graphical comparison of the $Bel$ values derived from the fusion of Set-$A$ and Set-$B$ evidence, while Figure \ref{BEL-PL-COMPARECVSS40-and-31}(b) contrasts the $Pl$ scores obtained from fusing evidence from both sets.

The $Bel$ (belief) scores and their corresponding ranks for evidence Set-$A$, which integrates data criticalities and security vulnerabilities using CVSS 4.0 metrics, are higher than those for evidence Set-$B$, which uses CVSS 3.1 metrics. This indicates stronger support for the evidence in Set-$A$. The $Pl$ (plausibility) scores for both evidence sets range from approximately 0.86 to 1.0, indicating high plausibility. However, the slightly higher overall $Pl$ scores for Set-$A$ further reinforce the conclusion that evidence in Set-$A$ is better supported.

The wider range of $Bel$ scores for Set-$B$ (0.1992 to 0.9834) and their subsequent ranks, compared to Set-$A$ (0.5465 to 0.9835), suggests greater variability or uncertainty in the strength of evidence for Set-$B$. The Dempster-Shafer theory is recognised for its nuanced representation of uncertainty, surpassing that of classical probability theory. The higher $Bel$ and $Pl$ scores for Set-$A$ indicate stronger support and lower uncertainty for this set of evidence.

Moreover, the top five ranking values based on sensor-generated data criticalities align with the rankings derived from the $Bel$ and $Pl$ values of the fused results from both Sets $A$ and $B$. However, the rankings based on security vulnerabilities using CVSS 4.0 and CVSS 3.1 metrics differ. This indicates that the fusion approach based on Dempster-Shafer (D-S) theory effectively integrates data criticality and security vulnerability.

Overall, the fusion results using D-S theory suggest that combining data criticalities and security vulnerabilities (evidence Set-$A$) is more strongly supported by the evidence compared to Set-$B$. The higher $Bel$ and $Pl$ scores for Set-$A$ indicate more robust support for this evidence set.

\subsection{Summary of Results}

In summary, the application of Dempster-Shafer (D-S) theory reveals that evidence Set-$A$, which integrates data criticalities and security vulnerabilities computed using CVSS 4.0 metrics, is better supported by the available data than evidence Set-$B$, which uses CVSS 3.1 metrics. The $Bel$ scores for Set-$A$ range from 0.5465 to 0.9835, indicating strong belief in the combined evidence, whereas $Bel$ scores for Set-$B$ vary more widely, from 0.1992 to 0.9834, suggesting greater uncertainty. Both sets have high $Pl$ scores (0.86 to 1.0), indicating high plausibility; however, Set-$A$ has slightly higher scores overall, reinforcing its stronger support. The D-S theory's quantitative approach provides a nuanced comparison of belief and plausibility, enhancing our understanding of the relative strengths of the two evidence sets.

The evaluation of critical IIoT sensor nodes and the comparative analysis highlight S07, S01, S04, S11, and S02 as highly critical sensors. Consequently, their integration into IIoT-enabled digital manufacturing systems necessitates prioritised security measures to ensure system protection. Critical sensors have a significant influence on production systems, particularly in maintaining high product quality. Compromising these critical IIoT devices can severely impact the expected outcomes of manufacturing processes. In this study, we focus on state-of-the-art digital wine manufacturing, where sensors S07, S01, S04, S11, and S02 correspond to the TSD-Sensor, FA-Sensor, RS-Sensor, ALC-Sensor, and VA-Sensor, respectively, all of which are crucial for maintaining wine quality. Therefore, these highly critical IIoT devices require enhanced security measures to effectively safeguard digital wine production systems.

\section{Conclusion}
\label{sec:conclusion}

This study presents a robust theoretical framework that bridges the Industrial Internet of Things (IIoT) domains, digital manufacturing, and cybersecurity. At its core, the framework integrates security vulnerability analysis and data criticality metrics to establish a comprehensive ranking of IIoT sensor node criticality. By combining these two dimensions, the proposed fusion-based methodology offers a more holistic risk assessment, capturing the potential impact of cyber threats on manufacturing processes. This innovative approach enables the identification of high-priority nodes for cybersecurity enhancement, allowing industries to optimise resource allocation to secure their IIoT ecosystems. The results validate the effectiveness of the fusion-based ranking in both strengthening sensor node security and preserving product quality. This validation reinforces the framework’s credibility and real-world applicability, making it a valuable tool for enhancing IIoT cybersecurity in manufacturing environments.

While this research represents a significant achievement, opportunities remain for further exploration and enhancement. Future work could expand the applicability of this approach and explore dynamic vulnerability assessment mechanisms for evolving threats.







\bibliographystyle{plain} 
\bibliography{cybersecurity}






\end{document}